\documentclass[pre,amsmath,twocolumn,showpacs]{revtex4}
\usepackage{bm}
\usepackage{amssymb}
\usepackage{graphicx}
\usepackage{subfigure}
\usepackage[dvips]{color}
\begin{document}

\title{The relation between the structure of blocked clusters and the relaxation dynamics in kinetically-constrained models}

\author{Eial Teomy}
\email{eialteom@tau.ac.il}
\author{Yair Shokef}
\email{shokef@tau.ac.il}

\affiliation{School of Mechanical Engineering, Tel Aviv University, Tel Aviv 69978, Israel}

\begin{abstract}

We investigate the relation between the cooperative length and the relaxation time, represented respectively by the culling time and the persistence time, in the Fredrickson-Andersen, Kob-Andersen and spiral kinetically-constrained models. By mapping the dynamics to diffusion of defects, we find a relation between the persistence time, $\tau_p$, which is the time until a particle moves for the first time, and the culling time, $\tau_c$, which is the minimal number of particles that need to move before a specific particle can move, $\tau_p=\tau^{\gamma}_c$, where $\gamma$ is model- and dimension dependent. We also show that the persistence function in the Kob-Andersen and Fredrickson-Andersen models decays subexponentially in time, $P(t)=exp\left[-\left(t/\tau\right)^\beta\right]$, but unlike previous works we find that the exponent $\beta$ appears to decay to $0$ as the particle density approaches $1$.
 
\end{abstract}


\maketitle

\section{Introduction}

Increasing the density of particles in granular materials causes them to undergo a transition from a fluid-like state, in which the particles can move relatively freely, to a jammed state, in which almost none of the particles can move~\cite{jamming,hecke}. In glasses, a similar transition occurs when the temperature is decreased \cite{glass,glass2}. As the material nears the glass or jamming transition, the system's relaxation time increases dramatically, until it diverges at the critical point \cite{angell}.

The various kinetically-constrained models \cite{review,review2,east,northeast,kronig} capture the essence of the glass or jamming transitions and there has been much recent activity on them. Some of these models simulate the way that particles block each other's movement by requiring that a particle can move only if its neighbors satisfy some condition \cite{sellitto,knights,DFOT,spiral2d,spiral2,spiral,spiral3d,jeng,elmatad}. Other models add driving forces which simulate the resistance of jammed systems to external forces \cite{fieldings,shokef,sellitto2,driving2,driving}. In general, the system is coarse-grained to a lattice, and each site is in one of two states, $0$ or $1$. In lattice-gas models a site in state $1$ represents a particle which may move to an adjacent vacant site, represented by state $0$, if its local neighborhood satisfies some model-dependent rule. In spin-facilitated models state $1$ represents a high density region in granular systems and an inactive
  region in glasses, while state $0$ represents either a low density region or an active region in granular matter and glass-forming liquids respectively. A site can change its state from $0$ to $1$ and vice versa, with a temperature-dependent rate if the site's local neighborhood satisfies some model-dependent rule.

In this paper we consider the Kob-Andersen (KA) \cite{ka} and Fredrickson-Andersen (FA) \cite{fa,fa2} kinetically-constrained models on one- and two-dimensional square lattices. In the FA spin-facilitated model, a site can change its state from $0$ to $1$ and vice versa if it has at least $m$ neighboring vacancies. In the KA lattice-gas model, a particle needs at least $m$ adjacent vacancies before and after the move in order to move to a nearest neighbor vacant site. Higher dimensional versions of these models with higher values of $m$ have also been investigated \cite{balogh,teomy2}.

The glass or jamming transitions result from cooperative dynamics, in the manner that particles are blocked by their neighbors, which in turn are blocked by their neighbors, and so on, such that in order for a single particle to move, many others need to move before it. The number of these ``shells" and their weight, represent the structural changes in the system as it nears the critical point, and they diverge at the critical point. In effect, they represent the minimal number of steps needed for a particle to move, which may be found by culling the shells iteratively. Above the critical density, or equivalently below the critical temperature, some of these shells cannot be culled since the particles in them block each other. This culling process is the usual manner to check whether a system is jammed or not, because if no shells remain after the culling then all the particles may move and the system is not jammed. The culling time represents a length scale related to relaxation of the system \cite{jeng,spiral3d}. We note here that this length scale is not the only way to quantify the relation between the structure of the system and its dynamics \cite{widmer1,widmer2,berthier,manning}, and it remains an open question which structural order parameter is a better choice.

In most previous works regarding the FA and KA models, the relaxation time was measured by the two-time density autocorrelation function \cite{nakanishi,schulz,einax,wyart,kuhlmann,leonard,mayer}. In this paper we use the persistence function, defined as the fraction of particles that have not yet moved until time $t$ (in lattice-gas models) or the fraction of sites that have not changed state until time $t$ (in spin-facilitated models). The persistence function was thoroughly investigated in the relatively simple $m=1$ models \cite{per1a,per1b,per1c,pergen,pergen2}, but there are also works on higher values of $m$ \cite{pergen,pergen2,per2,per3a,per3b}, and other kinetically-constrained models \cite{shokef}. Generally, the density autocorrelation function and the persistence function behave similarly.

In this paper we study the relation between the culling time and the relaxation time, obtained from the persistence function, and show that near the critical point the relation is a model-dependent power law which can be explained as a diffusion of rare droplets. We show that this is a general result by also considering another kinetically-constrained model, the spiral model \cite{spiral2d,spiral2}. In Section \ref{secmodels} we describe the models investigated in this paper. Our results for the culling time and the relaxation time are shown in Sections \ref{seccull} and \ref{secpers} respectively, and are compared in Section \ref{seccomp}. Section \ref{secsum} summarizes the paper.

\section{The Models}\label{secmodels}

We consider the KA and FA models on a $d$-dimensional square lattice. At time $t=0$, each site in the lattice is either in state $1$ with probability $\rho$, or in state $0$ with probability $v \equiv 1-\rho$ without correlations between sites. In this way, we probe the equilibrium distribution of the system. In the FA model, a site can change its state from $0$ to $1$ and vice versa if it has at least $m$ neighboring vacancies. In the KA model, sites at state $1$ are occupied by particles, and sites at state $0$ are vacant. A particle needs at least $m$ adjacent vacancies before and after the move in order to move to a nearest neighbor vacant site. We consider here three cases: $d=m=1$, $d=2,m=1$, and $d=m=2$. The first two cases $(m=1)$ in the KA model are equivalent to the simple symmetric exclusion principle (SSEP) model \cite{ssep1,ssep2}, in which a particle can move if it has a neighboring vacancy. When $m=1$ all the particles are able to move eventually (in the KA mod
 el) or change their state eventually (in the FA model), while if $m=2$ there is a system-size-dependent value of the density above which a finite fraction of the particles will not be able to move (KA model) \cite{toninelli} or change their state (FA model) \cite{fa}. In square systems of size $L\times L$, this critical vacancy density is given by \cite{holroyd}
\begin{align}
v_{c}=\frac{\lambda}{\ln L} ,\label{critv}
\end{align}
where $\lambda$ is a weak function of $L$ \cite{holroyd2,lambda}. In the system sizes we consider here, $\lambda\approx0.25$, whereas in the limit $L\rightarrow\infty$ it is equal to $\pi^{2}/18\approx0.55$.

We perform on these systems two types of dynamics: culling dynamics and real dynamics. In the culling dynamics, we iteratively remove the particles which are able to move (KA), or change to $0$ the state of the sites which are able to do so (FA). In the real dynamics, every time step $dt=1/N$, with $N$ being the number of sites, one of the sites is chosen randomly. 

In the FA model, in order for a site to change its state, it first must have $m$ neighboring vacancies as noted before. If this condition is satisfied, the site changes its state from $0$ to $1$ with probability $W_{01}$ and from $1$ to $0$ with probability $W_{10}$. In order to maintain detailed balance while maximizing the transition probabilities, we set
\begin{align}
&W_{01}=min\left(1,\frac{\rho}{v}\right) ,\nonumber\\
&W_{10}=min\left(1,\frac{v}{\rho}\right) .
\end{align}
In the KA model, if the chosen site is occupied, a random direction is also chosen, and the chosen particle can move in that direction if the neighboring site in that direction is empty, and the particle has at least $m$ neighboring vacancies before and after the move.

In the real dynamics, we use a continuous time, or rejection-free algorithm since at high densities the probability that an allowable move is randomly chosen is very small. In this algorithm we randomly generate the number of time steps that have passed between successive moves based on the probability that a move is possible. In this way, we do not wait for long periods of time until a move is made, but rather advance the clock in large random steps.

For the culling dynamics we define the culling time cumulative distribution $M^{(d,m)}(s)$ as the fraction of sites that started in state $1$ and didn't change to $0$ until iteration $s$ of the culling process, and for the real dynamics we define the persistence function $P^{(d,m)}(t)$ as the fraction of particles that have not yet moved (KA) or the fraction of sites that started from state $1$ and did not change to $0$ (FA) until time $t$. Obviously $M(0)=P(0)=1$ for all models. The culling time $\tau^{(d,m)}_{c}$ and the persistence time $\tau^{(d,m)}_{p}$, defined respectively as the average number of iterations needed to cull a particle and the average time until a particle moves (KA) or a site changes its state (FA) for the first time, are given by
\begin{align}
&\tau^{(d,m)}_{c}=\sum^{\infty}_{s=0}\frac{M^{(d,m)}\left(s\right)-M^{(d,m)}\left(\infty\right)}{1-M^{(d,m)}\left(\infty\right)} ,\nonumber\\
&\tau^{(d,m)}_{p}=\int^{\infty}_{0}\frac{P^{(d,m)}(t)-P^{(d,m)}\left(\infty\right)}{1-P^{(d,m)}\left(\infty\right)}dt , \label{taudef}
\end{align}
where $P^{(d,m)}(\infty)=M^{(d,m)}(\infty)=0$ if the system is unjammed, i.e. that all of the sites (particles) will be able to flip (move) eventually, and $P^{(d,m)}(\infty)\geq M^{(d,m)}(\infty)>0$ if the system is jammed, i.e. that some of the sites (particles) will never be able to flip (move). Therefore, $P^{(d,m)}(\infty)$ and $M^{(d,m)}(\infty)$ act as the system's Edwards-Anderson order parameter \cite{ea}. In the FA models, flipping sites only to $0$ may occur in the real dynamics, albeit with a negligible probability, and thus $M^{(d,m)}(\infty)=P^{(d,m)}(\infty)$. However, in the KA models it is possible that some particles will never be able to move but are still culled because other particles that may move but block them are culled, and thus $M^{(d,m)}(\infty)\leq P^{(d,m)}(\infty)$. Although the case $P^{(d,m)}(\infty)> M^{(d,m)}(\infty)=0$ is possible in finite systems, we assume that it does not occur in the thermodynamic limit since we encountered such a scen
 ario only in very small systems.

\section{Culling Dynamics}\label{seccull}

\subsection{Culling Dynamics for the $m=1$ FA and KA models}

In the $m=1$ models, there are no permanently frozen particles, and thus $M\left(\infty\right)=0$. Furthermore, we obtain an explicit expression for $M(s)$. The number of particles culled in the $s$'th step, $M(s-1)-M(s)$, is the number of particles that all of their $(s-1)$-nearest neighbors are occupied and at least one of the $s$-nearest neighbors is vacant. For $d=1$ this is
\begin{align}
M^{(1,1)}(s-1)-M^{(1,1)}(s)=\rho^{2(s-1)}\left(1-\rho^{2}\right) ,
\end{align}
and for $d=2$ it is
\begin{align}
M^{(2,1)}(s-1)-M^{(2,1)}(s)=\rho^{2s(s-1)}\left(1-\rho^{4s}\right) .
\end{align}
Solving these recursion equations yields for $d=1$
\begin{align}
M^{(1,1)}(s)=\rho^{2s} ,\label{m11}
\end{align}
and for $d=2$
\begin{align}
M^{(2,1)}(s)=\rho^{2s(s+1)} .\label{m21}
\end{align}
Hence, by substitution in Eq.~(\ref{taudef}) we find that the culling times are given by
\begin{align}
&\tau^{(1,1)}_{c}=\frac{1}{1-\rho^{2}} ,\nonumber\\
&\tau^{(2,1)}_{c}=\sum^{\infty}_{s=0}\rho^{2s(s+1)}=\frac{\Theta_2\left(0,\rho^{2}\right)}{2\sqrt{\rho}} ,
\end{align}
where $\Theta_{2}$ is the Jacobi Theta function \cite{jacobi}. At high particle densities, $v\ll1$, we may approximate $\tau^{(2,1)}_{c}$ by
\begin{align}
\tau^{(2,1)}_{c}\approx\sqrt{\frac{\pi}{8v}} .
\end{align}

In a similar manner for general dimensions, by calculating the number of $s$-nearest neighbors in a $d$-dimensional hypercubic lattice, $G_{d}(s)$, we find that
\begin{align}
M^{(d,1)}(s)=\rho^{G_{d}(s)-1} . \label{eqmd}
\end{align}
It was shown in \cite{sloane} that $G_{d}(s)$ is a polynomial given by
\begin{align}
G_{d}(s)=\sum^{d}_{k=0}\left(\begin{array}{c}d\\k\end{array}\right)\left(\begin{array}{c}s-k+d\\d\end{array}\right) . \label{eqmd2}
\end{align}

At high particle densities, $v=1-\rho\ll1$, we can find an approximation for $\tau_{c}$ in any dimension. From Eq. (\ref{eqmd}) we find that $\tau^{(d,1)}_{c}$ is given by
\begin{align}
\tau^{(d,1)}_{c}=\sum^{\infty}_{s=0}\rho^{G_{d}(s)-1}=\sum^{\infty}_{s=0}\exp\left[\ln\rho\left(G_{d}(s)-1\right)\right] .
\end{align}
We now note that $G_{d}(s)$ is a polynomial of order $d$ with the coefficient of $s^{d}$ given by
\begin{align}
G_{d}(s)=\sum^{d}_{k=0}\left(\begin{array}{c}d\\ k\end{array}\right)\frac{1}{d!}s^{d}+O(s^{d-1})=\frac{\left(2s\right)^{d}}{d!}+O(s^{d-2}) .
\end{align}
Changing the sum over $s$ to an integral over $x=sv^{1/d}$ yields
\begin{align}
&\tau^{(d,1)}_{c}\approx v^{-1/d}\int^{\infty}_{x=0}\exp\left[-\frac{\left(2x\right)^{d}}{d!}\right]dx=\nonumber\\
&=\frac{\Gamma\left(1+\frac{1}{d}\right)\left(d!\right)^{1/d}}{2v^{1/d}} .\label{taucapp}
\end{align}

Except for the non-trivial prefactor, the dependence of $\tau_{c}$ on the vacancy density $v$ comes simply from the fact that $\tau_{c}$ is the distance to the nearest vacancy, which scales as $v^{-1/d}$.

\subsection{Culling Dynamics for the $m=2$ FA and KA models}

In the $d=m=2$ models, we find $M^{(2,2)}(s)$ and $\tau^{(2,2)}_{c}$ numerically by running simulations on square systems of size $L\times L$, with $L=100$ or $1000$. We only consider densities below the critical density (see Eq. (\ref{critv})), $\rho_{c}(L=100)\approx0.94$ and $\rho_{c}(L=1000)\approx0.96$, since in the thermodynamic limit the critical density is $\rho_c\left(L=\infty\right)=1$ and thus the results relevant to this limit are below the size-dependent critical density.

\begin{figure}
\includegraphics[width=0.45\columnwidth]{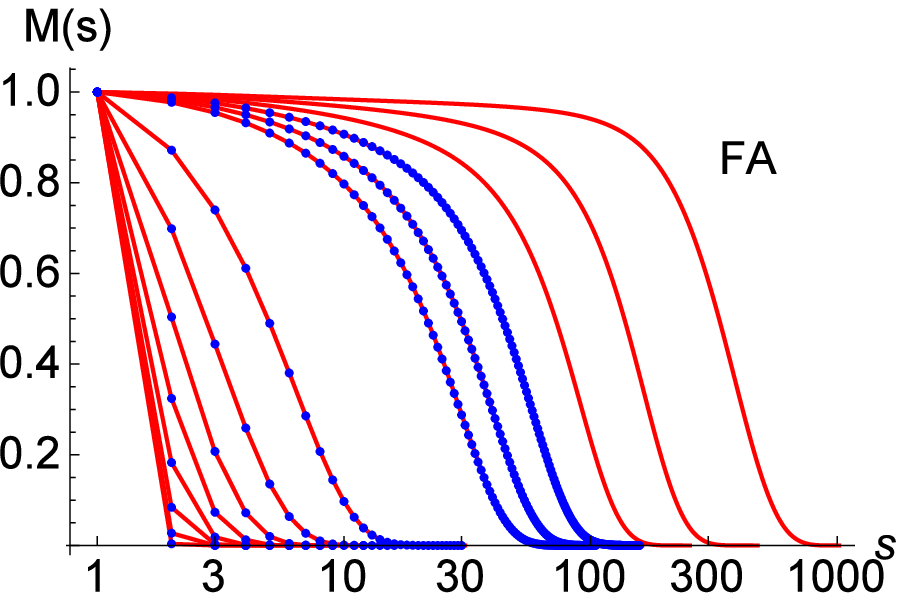}
\includegraphics[width=0.45\columnwidth]{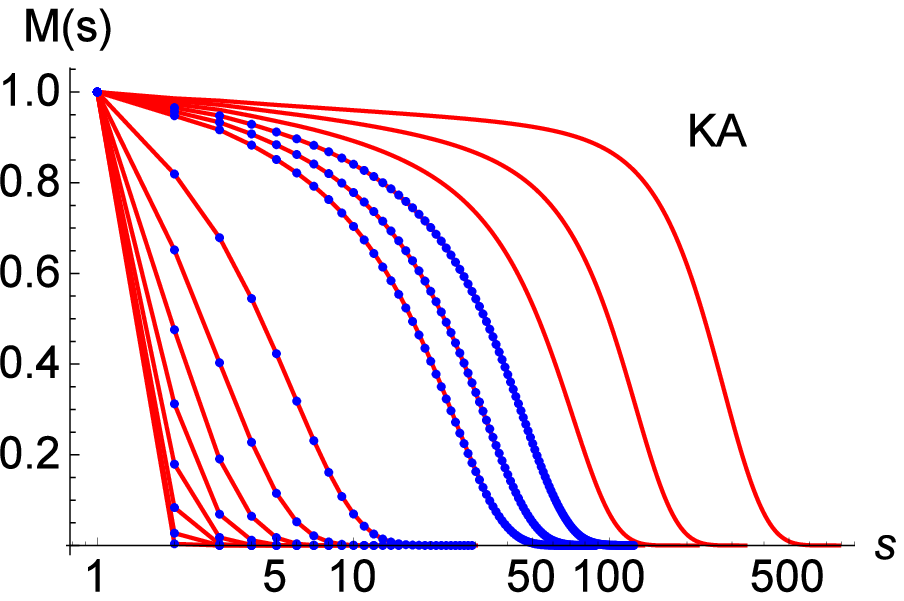}\\
\includegraphics[width=0.45\columnwidth]{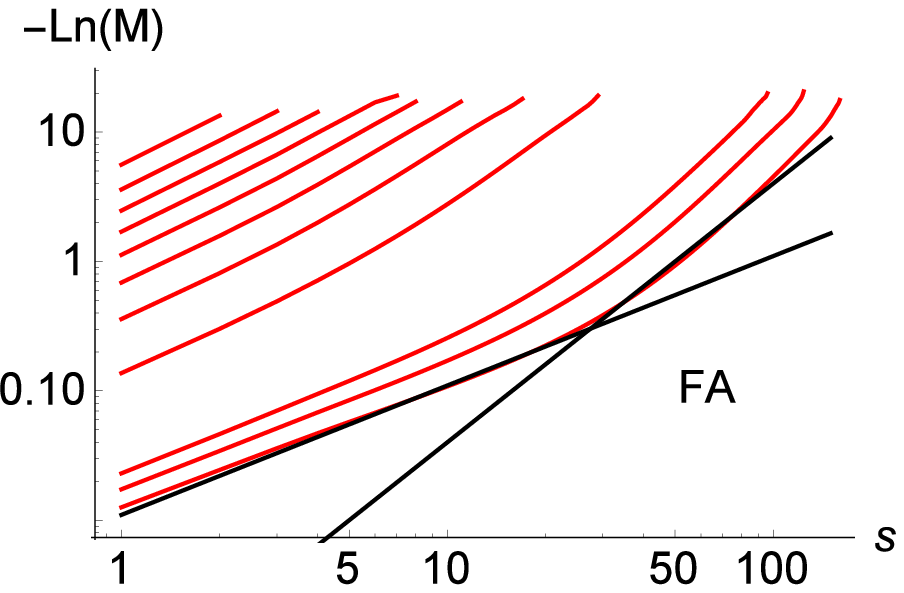}
\includegraphics[width=0.45\columnwidth]{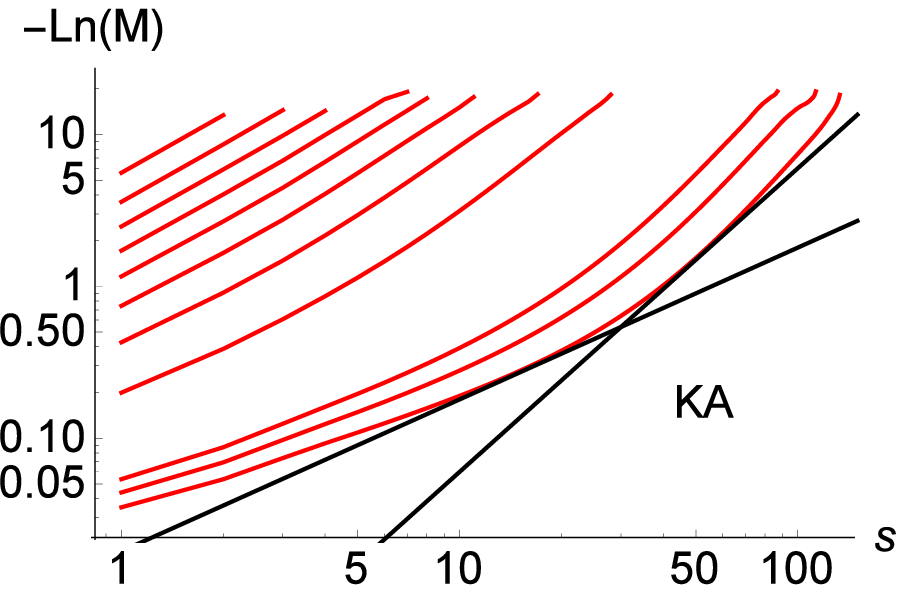}
\caption{The fraction of sites that haven't yet been culled, $M$, (top panels) and its logarithm (bottom panels) as a function of the iteration number $s$ for the $d=m=2$ FA model (left panels) and KA model (right panels). The system's linear size is either $L=100$ (blue dots) or $L=1000$ (red continuous lines). Each curve is for a different density (from left to right) $\rho=0.1,0.2,0.3,0.4,0.5,0.6,0.7,0.8,0.9,0.91,0.92,0.93,0.94,0.95$. There is almost no difference between the different system sizes. The bottom panels show only data for $L=1000$. The straight black lines in the bottom panels are $\sim s$ and $\sim s^{2}$. There is almost no difference between the FA and the KA models.}
\label{m_v_s}
\end{figure}

\begin{figure}
\includegraphics[width=\columnwidth]{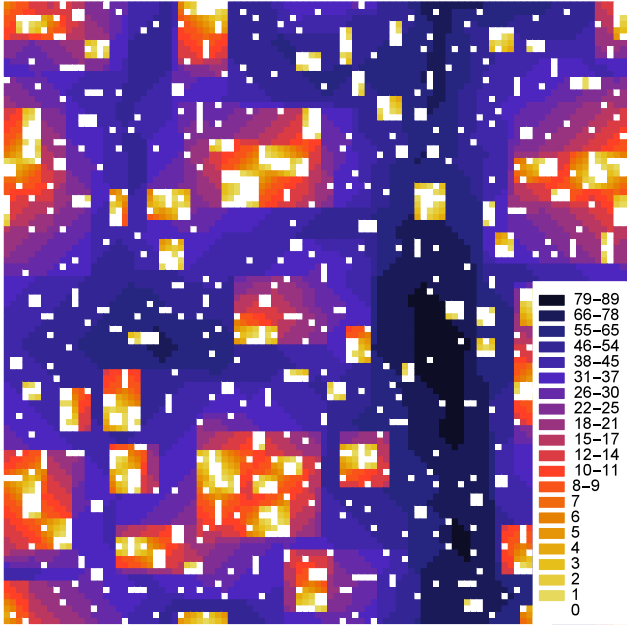}
\caption{The culling time $s$ for each site in a typical $100\times100$ configuration in the FA model at $\rho=0.92$. The legend shows the range of $s$ represented by each color. The diagonal borders between regions of different colors, which are indicative of a two-dimensional culling process, are clearly seen at large scales. See for example the diagonals at the lower left corner, and in the region above the legend to the right.}
\label{logcull}
\end{figure}

Figure \ref{m_v_s} shows the dependence of $M(s)$ on $s$. At small $s$ we find an exponential decay $-\ln M^{(2,2)}\sim s$, which is similar to the behavior of $M^{(1,1)}$ in Eq. (\ref{m11}), while for large $s$, the form is Gaussian $-\ln M^{(2,2)}\sim s^{2}$, which is similar to $M^{(2,1)}$ in Eq. (\ref{m21}). The reason is that for small $s$, the particles are culled mostly one by one such that the behavior is quasi-one-dimensional, and when the empty region is large enough, the particles around it are culled by diagonal shells as a two-dimensional system, see Fig. \ref{logcull}.

\begin{figure}
\includegraphics[width=\columnwidth]{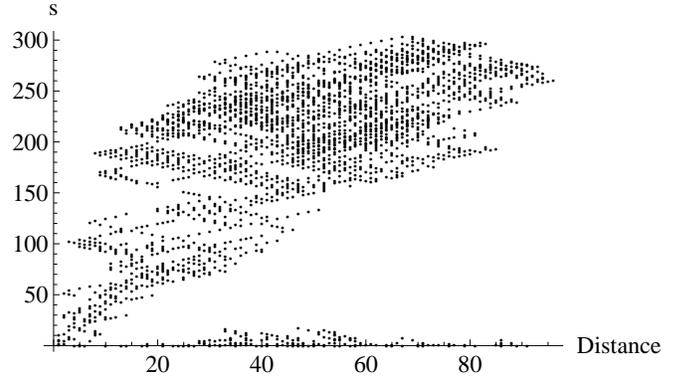}
\caption{A scatter plot of the culling time $s$ vs. the distance $(\left|\Delta x\right|+\left|\Delta y\right|)$ to the nearest critical droplet for a single $100\times100$ configuration at the critical density $\rho=0.951$ for the d=m=2 FA model. Lines with positive (negative) slope indicate culling around the empty region away (toward) the seed of the droplets, see text. At intermediate times $(50\leq s\leq150)$ the culling time generally grows with the distance from the droplet, since at those times the droplet in the configuration shown here expanded away from the seed in two of the four directions. At longer times, the droplet expanded in all directions, including toward the seed. The points with low $s$ and large distance represent small, active regions which are far from the critical droplet.}
\label{cull_v_dis}
\end{figure}

\begin{figure}
\includegraphics[width=\columnwidth]{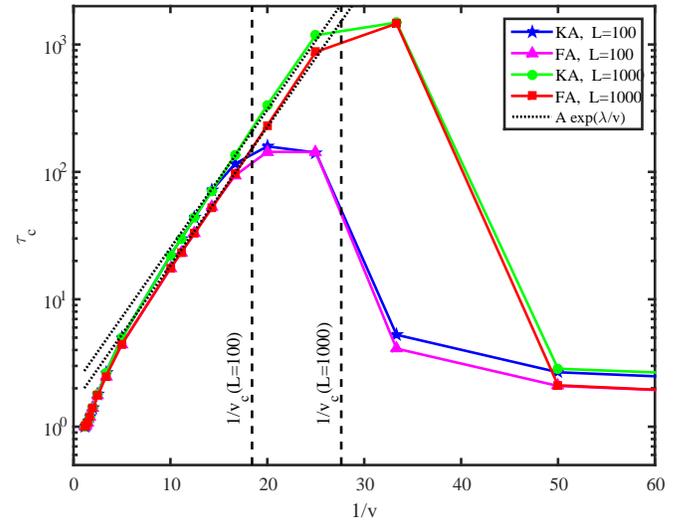}
\caption{The mean culling time $\tau_{c}$ vs. the reciprocal of the vacancy density $1/v$ for $d=m=2$. There is almost no difference between the $KA$ and the $FA$ models. Below the jamming density, there is almost no difference between $L=100$ and $L=1000$, and the mean culling time may be fitted to $\tau_{c}= A \exp\left(\lambda/v\right)$ with $A\approx 2.1$(KA),$1.5$(FA), and $\lambda=0.25$, as shown by the black dotted lines. $\tau_c$ reaches a maximum around the critical density (vertical dashed black lines), because at higher densities the frozen sites can no longer be culled and thus do not contribute to $\tau_c$.}
\label{tc_v_r}
\end{figure}

In the $m=1$ models, the culling process is equivalent to expanding the vacant regions, and thus the culling time for a given site is its distance to the nearest vacancy, $\tau_{c}\sim v^{-1/d}$, see Eq. (\ref{taucapp}) above. In the $m=2$ models, the culling process is dominated by critical droplets \cite{holroyd}, which are small regions that may be expanded by the culling process to include the entire system. Hence, the culling time is the distance to the nearest seed for a critical droplet, if the site is far enough from the droplet, as shown in Fig. \ref{cull_v_dis}. Because the probability of a given site to seed a critical droplet is $\exp\left(-2\lambda/v\right)$, with $\lambda\approx0.25$ in the sampled density range \cite{lambda}, the average distance from a droplet, and thus the mean culling time, should scale as $\exp\left(\lambda/v\right)$. We see from Fig. \ref{tc_v_r} that this form of the scaling is consistent with our numerical data.

\section{Persistence in the physical dynamics}\label{secpers}

\subsection{Real Dynamics for the $m=1$ FA and KA models}

Since the $m=1$ KA lattice gas is equivalent to SSEP, instead of considering motion of particles, we may think of the dynamics as diffusion of vacancies, such that the persistence of a given site is the mean first passage time of vacancies to that site. Since the vacancies can diffuse freely, all particles will eventually move and thus $P(\infty)=0$. At high particle density, $v\ll1$, we may make the approximation that the vacancies are independent, and allow two (or more) vacancies to occupy the same site. Furthermore, at long times the discrete nature of the lattice becomes irrelevant and we may use results from continuous models. Under these approximations, the long time behavior of the mean first passage time distribution, and thus of the persistence function, is described by \cite{bramson,diffusion,meerson} 
\begin{align}
P^{(d,1)}(t)=\left\{\begin{array}{lr}
\exp\left[-2v\sqrt{\frac{Dt}{\pi}}\right]&d=1\\\\
\exp\left[-\frac{4\pi Dvt}{\ln Dt/R^{2}}\right]&d=2\\\\
\exp\left[-\left(d-2\right)S_{d}R^{d-2}Dvt\right]&d\geq3
\end{array}\right. , \label{pt}
\end{align}
where $S_{d}=2\pi^{d/2}/\Gamma\left(d/2\right)$ is the surface area of the $d$-dimensional unit sphere, and $D$ is the self-diffusion coefficient for the motion of the vacancies. For the $m=1$ models, $D=1/(2d)$. In continuous models, $R$ is the radius of the trapping region. 
In a discrete lattice, in which each site contains at most one particle, $R=O(1)$. 

At short times we may use a mean-field approximation, such that the probability that a particle can move to an adjacent site (for the first time, since this is an approximation for short times) is $v$, and thus 
\begin{align}
\frac{\partial P}{\partial t}=-vP(t) ,
\end{align}
which yields $P(t)=e^{-vt}$. Note that in kinetically constrained models, by construction the occupation probabilities of neighboring sites at a given time are uncorrelated, the dynamics are spatially heterogeneous~\cite{review2,wyart}. In the appendix we derive an exact expression for the one-dimensional case at all times, under the approximation that the diffusing vacancies are independent. 

The FA Ising model may be thought of as a diffusion-reaction model, which behaves similarly. At high particle densities in the FA model, $W_{10} \ll W_{01}$, and sites in state $0$ can be considered to change practically instantly to state $1$ (when the kinetic constraint does not prevent them from doing so) compared to the time it takes a site in state $1$ to flip. Thus, when a state $1$ flips and immediately after that its state $0$ neighbor flips, it appears as if the state $1$ moved. This effective movement happens on a different time scale than in the KA model, because the rate $W_{10}$ is smaller than $1$, and thus time should be normalized by $W_{01}W_{10}=min\left(\frac{\rho}{v},\frac{v}{\rho}\right)$ in order to map the FA dynamics on those of the KA model. In what follows we thus normalize time by $W_{01}W_{10}$ and interpret the rates $W_{10}$ and $W_{01}$ as equal to unity in the KA model.

\begin{figure}[htb]
\includegraphics[width=0.9\columnwidth]{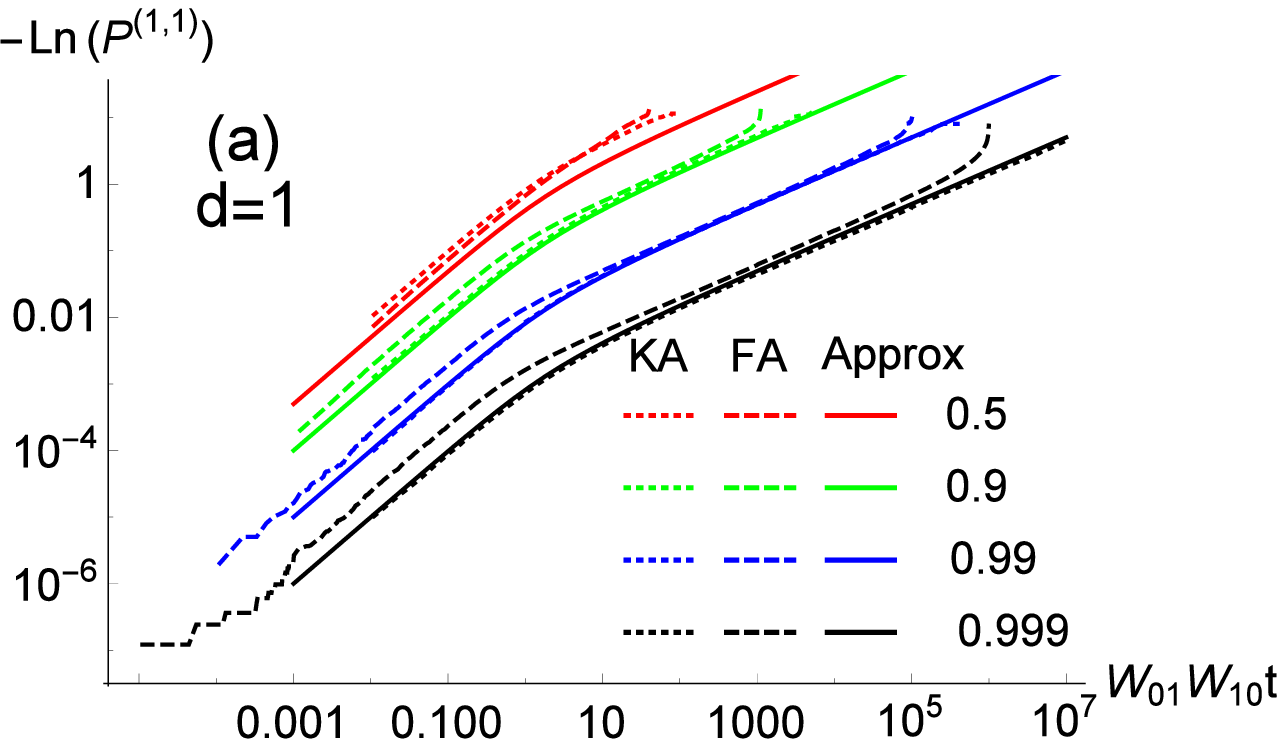}\\
\includegraphics[width=0.9\columnwidth]{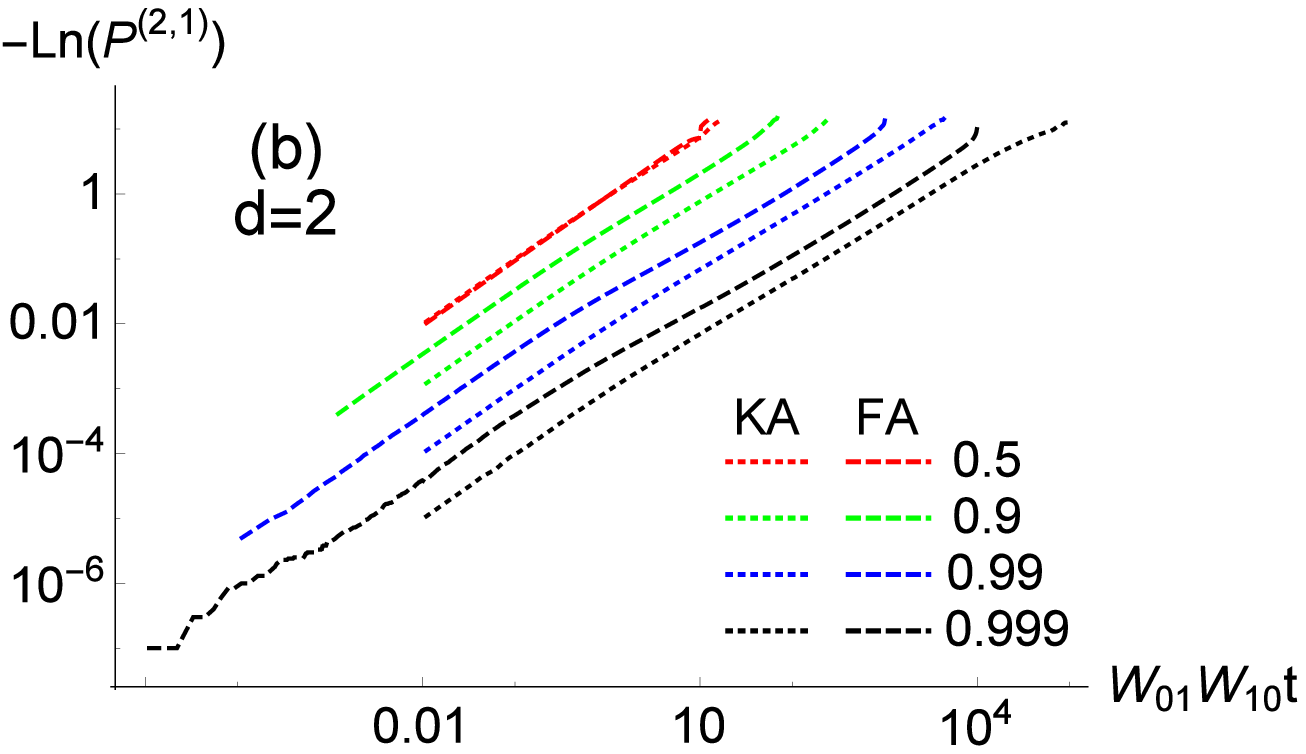}\\
\includegraphics[width=0.9\columnwidth]{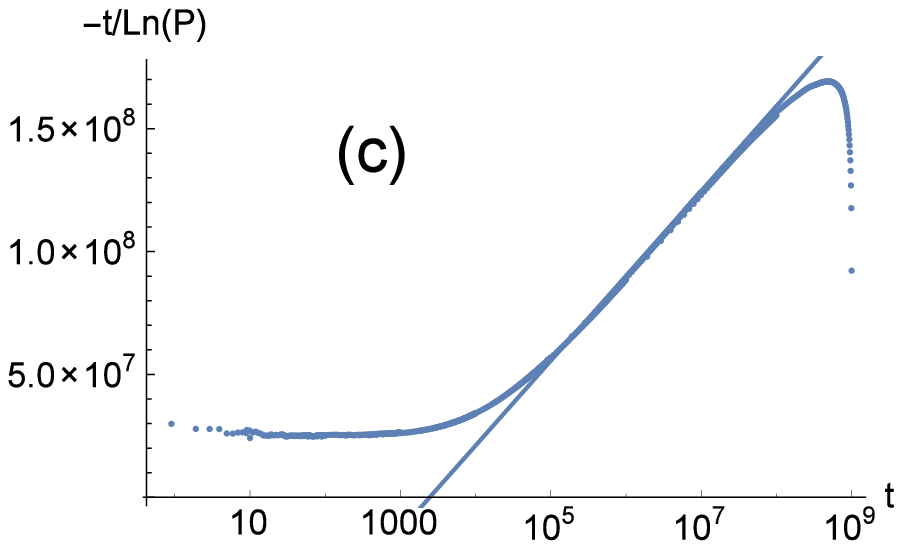}
\caption{The persistence function $P$ as a function of the normalized time $W_{01}W_{10}t$ in the $m=1$ models in $d=1$ (a) and $d=2$ (b) for the FA (dashed lines) and KA (dotted lines) models. In panel (a), the continuous lines are the analytical approximation, Eq. (\ref{pt}). As the scaling from Eq. (\ref{pt}) for $d=2$ is difficult to see from panel (b), we show in panel (c) $t/\ln P$ vs. $t$ for $\rho=0.9999$ in the $d=2$ FA model. At small times $P$ decreases exponentially with $t$, but at longer times it decreases exponentially with $t/\ln t$ as expected from Eq. (\ref{pt}). The drop at long times is due to finite-size effects, and the thin line is $\sim\ln t$.}
\label{pvtm1}
\end{figure}

Figure \ref{pvtm1} shows the persistence function for the $m=1$ models. We see that the FA and KA models behave similarly, except for a prefactor, and that the analytical approximation, Eq. (\ref{pt}), is in good agreement with the numerical results.

\subsection{Real Dynamics for the $m=2$ FA and KA models}

\begin{figure}
\includegraphics[width=0.9\columnwidth]{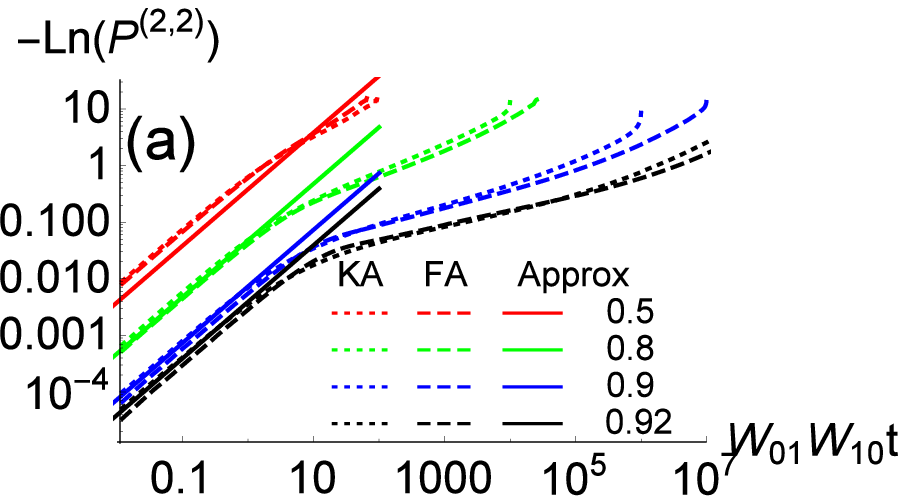}\\
\includegraphics[width=0.9\columnwidth]{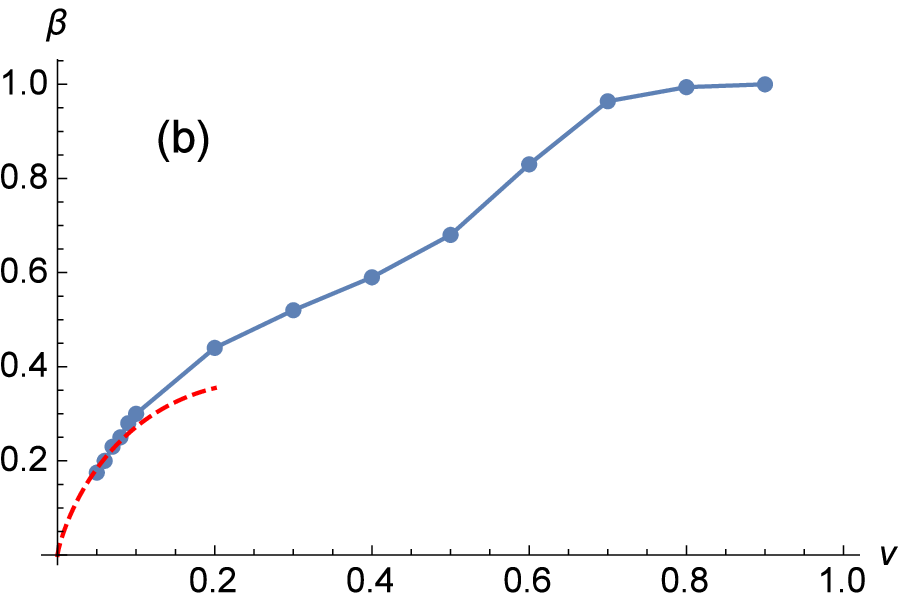}
\caption{(a) The persistence function $P$ as a function of the normalized time in the $m=2$ FA (dashed lines) and KA (dotted lines) models.
At short times the exponential decay of the persistence function agrees with Eqs.~(\ref{pt1},\ref{omega}), denoted by the solid lines, but at longer times it decays subexponentially as $-ln P\sim t^{\beta}$. (b) The exponent $\beta$ vs. the vacancy density $v$. The blue symbols are the numerical results, and the dashed red line is the analytical approximation for small $v$, Eq. (\ref{betavsv}).}
\label{pvtm2}
\end{figure}

Similarly to the $m=1$ models, at high densities the $m=2$ FA model behaves as the $m=2$ KA model under the proper time normalization for the exact same reasons as in the $m=1$ models. At short times, the persistence decays exponentially as
\begin{align}
P(t)=e^{-\Omega W_{10}t} ,\label{pt1}
\end{align}
where in $d=2$,
\begin{align}
\Omega=\left(1-\rho^{3}\right)^{2}v \label{omega}
\end{align}
is the probability that a random particle has at least two neighboring vacancies before and after the move. This result is obtained from a mean-field approximation, valid for short times, similarly to the analysis above for the $m=1$ models. This exponential decay continues until time $1/W_{10}$, which is the typical time at which all the sites that were able to flip at $t=0$ have flipped (FA) or all the particles that were able to move at $t=0$ have moved (KA). After that time, we see from the numerical results that the persistence decays as a stretched exponential
\begin{align}
P(t)=e^{-\Omega\left(W_{10}t\right)^{\beta}} ,\label{pt2}
\end{align}
as shown in Fig.~\ref{pvtm2}. Note however that in the thermodynamic limit and at extremely long times the persistence function eventually decays exponentially~\cite{pergen}.

Combining Eqs. (\ref{taudef}), (\ref{pt1}) and (\ref{pt2}) we find that the persistence time may be approximated by
\begin{align}
W_{10}\tau_{p}=\frac{1-e^{-\Omega}}{\Omega}+\frac{\Gamma\left(\frac{1}{\beta},\Omega\right)}{\beta \Omega^{1/\beta}} ,
\end{align}
where $\Gamma\left(a,z\right)$ is the incomplete Gamma function. In the limit of small $v$ (and thus small $\beta$), $\tau_{p}$ may be further approximated by
\begin{align}
W_{10}\tau_{p}\approx\sqrt{\frac{2\pi}{\beta}}\left(\frac{1}{9ev^{3}\beta}\right)^{1/\beta} .\label{beta}
\end{align}

Previous numerical studies \cite{perm2} have shown that in a $50\times50$ system, the exponent $\beta$ converges to a value of $0.42$ as the density is raised to $\rho=0.917$, near the critical density for jamming in a system of that size, $\rho_{c}\approx0.94$ \cite{teomy}, while we get $\beta=0.25$ at $\rho=0.92$ and it clearly does not converge. The reason for this apparent discrepancy lies in the preparation protocol. In our simulations, each site at time $t=0$ is in state $1$ with probability $\rho$ and in state $0$ with probability $v$, and thus the initial configuration is chosen from the equilibrium distribution. In the simulations reported in \cite{perm2}, all the spins were initially set to $0$, the system was evolved for a long time until it apparently reached equilibrium, and then the measurement of the persistence function started. However, we suspect that these simulations did not equilibrate. Indeed, by simulating a $50\times50$ system at $\rho=0.92$ with such 
 a quenched initial condition, we find $\beta=0.42$ if we wait for $10^5$ steps per site, while by waiting for $10^8$ steps per site we get $\beta=0.32$, still far from the equilibrium result. It would be interesting to test the convergence of $\beta$ to its equilibrium value by waiting considerably longer times after such quenches.

\section{Comparison between the culling dynamics and the real dynamics}\label{seccomp}

In order to compare the culling time $\tau_{c}$ and the persistence time $\tau_{p}$, we return to the picture of diffusing vacancies. For $m=1$, combining Eqs. (\ref{taudef}),(\ref{taucapp}) and (\ref{pt}) for $d=1$ and $d\geq3$ yields
\begin{align}
\tau^{(d,1)}_{p}=\left\{\begin{array}{lr}
\pi\tau^{2}_{c}&d=1\\\\
\tilde{c}_{d}\tau^{d}_{c}&d\geq3
\end{array}\right. ,\label{reld1}
\end{align}
where $\tilde{c}_{d}$ is some constant.
For $d=2$ the integral
\begin{align}
\tau^{(2,1)}_{p}=\int^{\infty}_{1}P^{(2,1)}(t)dt ,
\end{align}
where $P^{(2,1)}(t)$ is given by Eq. (\ref{pt}), cannot be computed exactly for any finite $v$, and requires more work to find the asymptotic expansion for small $v$. We first change the integration variable from $t$ to $x=4\pi Dvt$
\begin{align}
\tau^{(2,1)}_{p}=\frac{1}{4\pi Dv}\int^{\infty}_{4\pi Dv}\exp\left[-\frac{x}{\ln (x/c_{2}v)}\right]dx ,
\end{align}
where
\begin{align}
c_{2}=4\pi R^{2} ,
\end{align}
and as noted above $R=O(1)$. 
We now divide the range of integration to three parts: $4\pi Dv$ to $1$, $1$ to $1/(c_{2} v)$, and $1/(c_{2}v)$ to $\infty$. The first part is negligible because its total contribution is smaller than $1$. The third part is negligible because in the limit of $v\rightarrow0$, both the integrand and the range of integration go to $0$. Hence,
\begin{align}
\tau^{(2,1)}_{p}\approx\frac{1}{4\pi Dv}\int^{1/(c_{2}v)}_{1}\exp\left[-\frac{x}{\ln x-\ln (c_{2}v)}\right]dx .
\end{align}
In this region, $\left|\ln x\right|<\left|\ln c_{2}v\right|$, and thus we may further approximate $\tau^{(2,1)}_{p}$ by
\begin{align}
&\tau^{(2,1)}_{p}\approx\frac{1}{4\pi Dv}\int^{1/(c_{2}v)}_{1}\exp\left[\frac{x}{\ln (c_{2}v)}\right]dx\approx\nonumber\\
&\approx-\frac{\ln(c_{2}v)}{4\pi Dv} ,\label{tau2d}
\end{align}
where in the last approximation we used $v\ll1$. Combining Eq. (\ref{taucapp}) and (\ref{tau2d}) yields
\begin{align}
\tau^{(2,1)}_{p}\sim\tau^{2}_{c}\ln\tau_{c} .\label{reld2}
\end{align}

In the $m=2$ models, the dynamics is dominated by movement of droplets, not of individual vacancies. We recall that the droplets appear with an effective density of
\begin{align}
\tilde{v}=\exp\left(-2\lambda/v\right) ,
\end{align}
and that $\tau_{c}\sim\exp\left(\lambda/v\right)\sim\tilde{v}^{-1/2}$. The self-diffusion coefficient of particles in the $m=2$ model is given by~\cite{toniphd}
\begin{align}
D=\exp\left(-2\lambda/v\right)\sim\tau^{-2}_{c} .\label{diffm2}
\end{align}
We are interested in the self diffusion of droplets. The particles inside the droplets are the most mobile particles in the system, and thus contribute the most to the self diffusion coefficient of particles in the system. Therefore, the self diffusion coefficient for the droplets may be approximated by the self diffusion coefficient of the particles, given by Eq.~(\ref{diffm2}).
Using Eq. (\ref{diffm2}) and changing $v$ to $\tilde{v}$ in Eq. (\ref{tau2d}) yields
\begin{align}
\tau^{(2,2)}_{p}\sim\tau^{4}_{c}\ln\tau_{c} .\label{relm2}
\end{align}

From this relation we can find an approximation for $\beta(v)$ at small $v$. Combining Eqs. (\ref{beta}), (\ref{diffm2}) and (\ref{relm2}) for the KA model yields
\begin{align}
\sqrt{\frac{2\pi}{\beta}}\left(\frac{1}{9ev^{3}\beta}\right)^{1/\beta}=\frac{\lambda}{2\pi v}\exp\left(4\lambda/v\right) .
\end{align}
Solving for $v$ yields
\begin{align}
v=\frac{4\beta\lambda}{\left(\beta-3\right)W\left[\frac{\beta}{\beta-3}\left(576e\lambda^{3}\beta\left(\frac{\beta}{2\pi c^{2}_{2}}\right)^{\beta/2}\right)^{1/\left(3-\beta\right)}\right]} ,\label{vvsb}
\end{align}
where $W\left[z\right]$ is the product-log function~\cite{prod_log} defined as the solution to
\begin{align}
z=W\left[z\right]\exp\left(W\left[z\right]\right) .
\end{align}
Expanding Eq. (\ref{vvsb}) for small $\beta$ yields
\begin{align}
v=\frac{3\lambda\beta}{2\ln\left(1/\beta\right)} .
\end{align}
Solving for $\beta$, and approximating for small $v$ yields
\begin{align}
\beta=\frac{2v}{3\lambda}\ln\left(\frac{3\lambda}{2v}\right) ,\label{betavsv}
\end{align}
which for small $v$ is shown on Fig.~\ref{pvtm2}b to roughly agree with the value of $\beta$ obtained from the stretched exponential form of the persistence function, Eq.~(\ref{pt2}).

\begin{figure}
\includegraphics[width=\columnwidth]{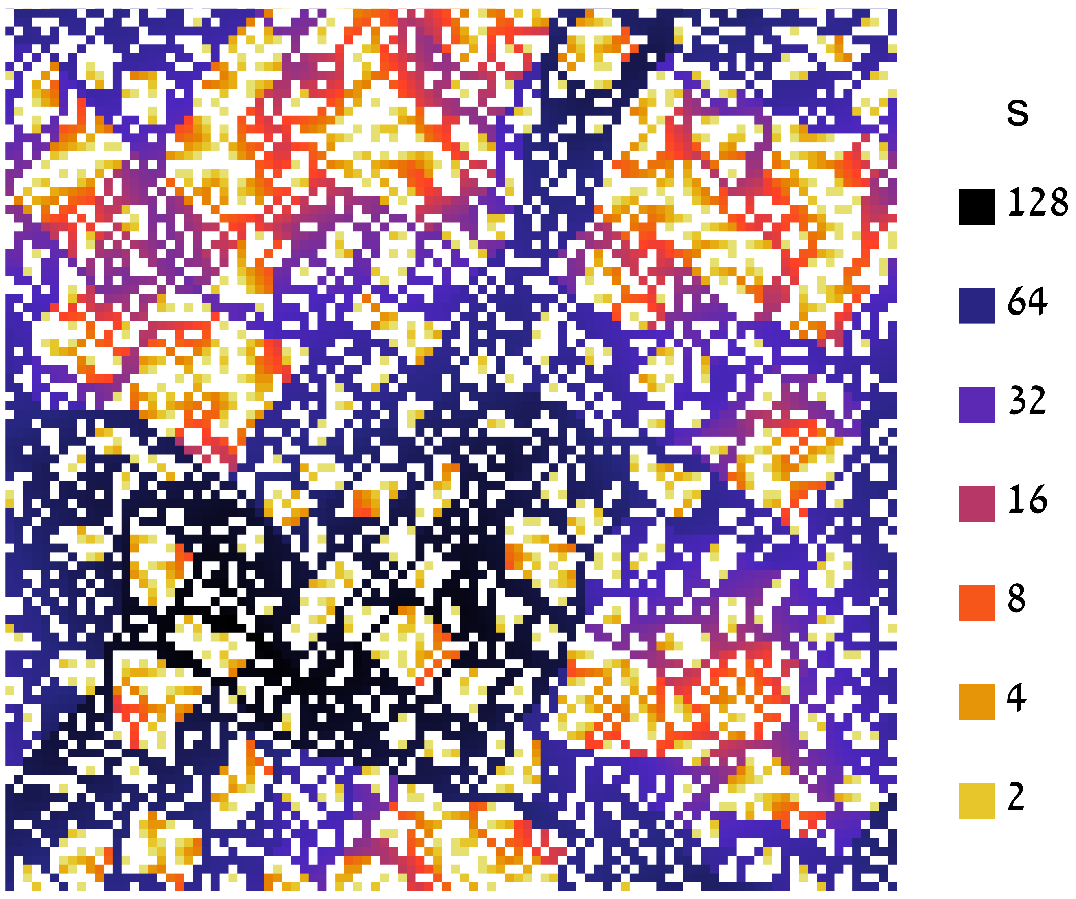}
\includegraphics[width=\columnwidth]{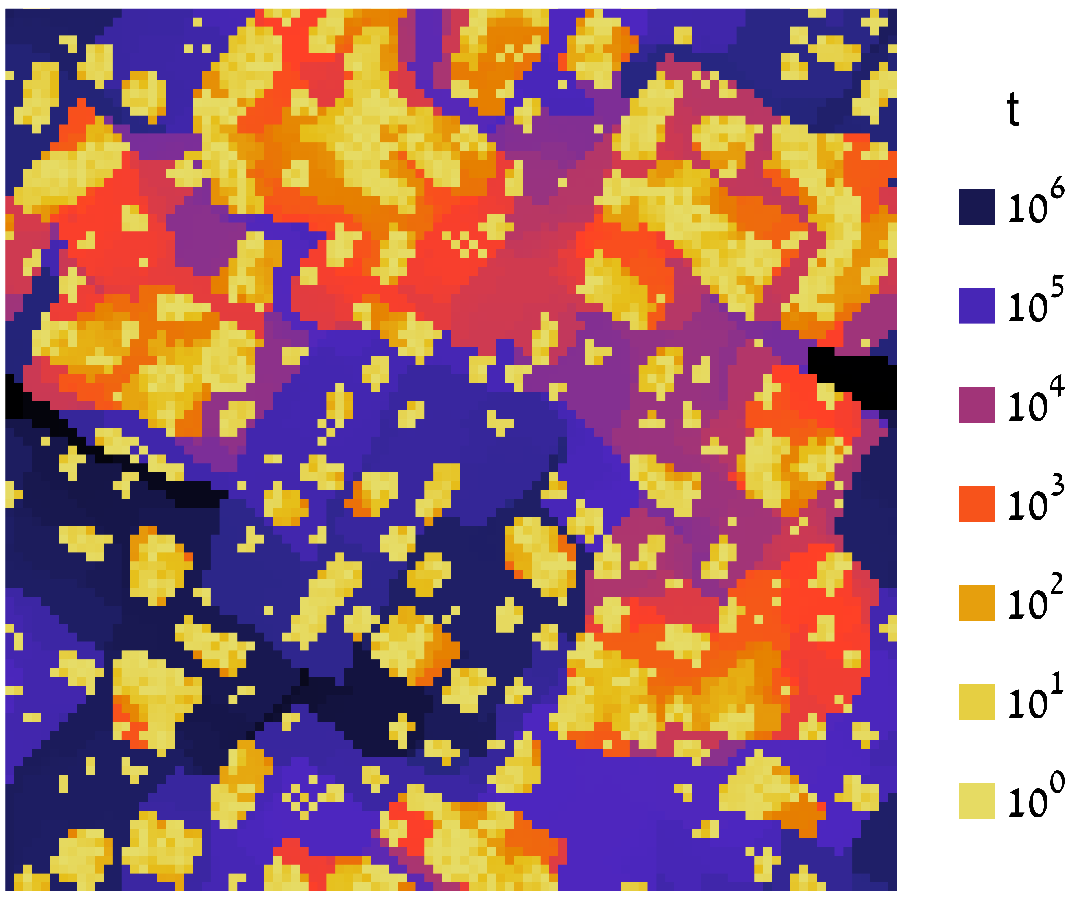}
\caption{Spatial structure of the culling (top) and persistence (bottom) times in the spiral model at $\rho=0.61$, which is slightly below the critical density for the simulated system size of $L=100$. The one-dimensional structures running along the diagonal directions of the lattice are clearly seen.}
\label{spiral_map}
\end{figure}

\begin{figure}
\includegraphics[width=\columnwidth]{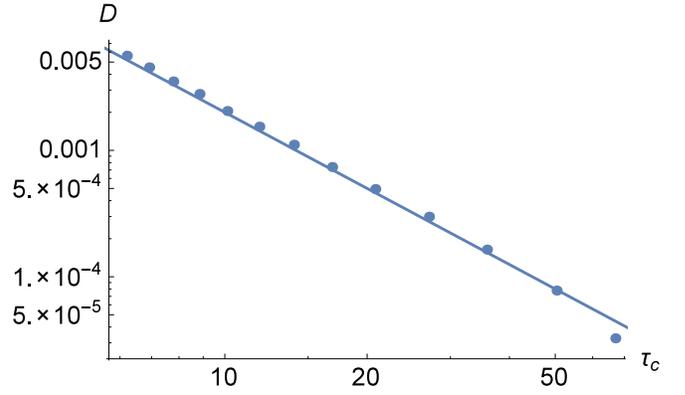}
\caption{The self-diffusion coefficient $D$ in the two-dimensional spiral model vs. the culling time $\tau_c$. The continuous line is $\sim\tau^{-2}_{c}$.}
\label{spiraldiffusion}
\end{figure}

In order to show that the relation between $\tau_{p}$ and $\tau_{c}$ is general, we also consider here the two-dimensional spiral model~\cite{spiral2d,spiral2}. This model jams at a finite density, at which the frozen structures are one-dimensional strings that run along the diagonal directions of the lattice. Below the critical density, but near it, the largest contribution to the persistence time comes from such almost-frozen strings, and thus we may approximate this as a quasi-one-dimensional process (see Fig.~\ref{spiral_map}), which leads to
\begin{align}
\tau^{spiral}_p\sim \tau^{2}_{c}/D .
\end{align}
As the density approaches the critical density, we see from the numerical results shown in Fig. \ref{spiraldiffusion}, that the diffusion coefficient approaches zero as
\begin{align}
D\sim\tau^{-2}_{c} ,
\end{align}
and thus
\begin{align}
\tau^{spiral}_p\sim\tau^{4}_{c} .\label{relspiral}
\end{align}

\begin{figure}
\includegraphics[width=\columnwidth]{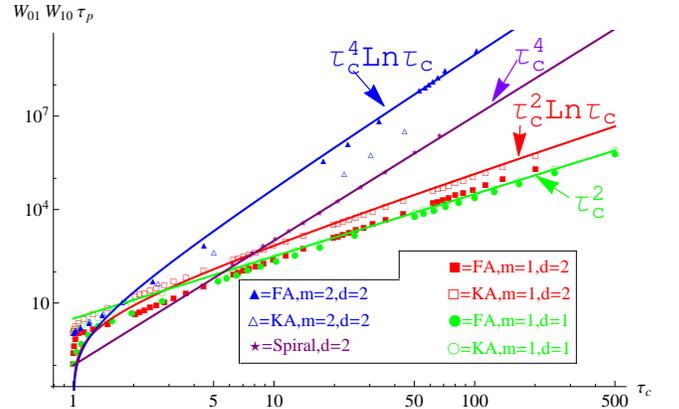}
\caption{The relation between the normalized persistence time $W_{10}W_{01}\tau_{p}$ and the culling time $\tau_{c}$ for the FA and KA models and for the spiral model. The continuous lines are the analytical approximations, Eqs. (\ref{reld1}), (\ref{reld2}), (\ref{relm2}), and (\ref{relspiral}).}
\label{p_v_c}
\end{figure}

Figure \ref{p_v_c} shows the excellent agreement between the numerical results and Eqs. (\ref{reld1}), (\ref{reld2}), (\ref{relm2}) and (\ref{relspiral}) at densities slightly below the critical density ($\rho_c=1$ for the FA and KA models, and $\rho_c\approx0.7$ for the spiral model). It would be interesting to study the structural connections between the directed percolation underlying the jammed structures in the spiral model and the one-dimensional nature of the relaxation processes in it. Furthermore, it would be interesting to identify possible logarithmic corrections to Eq.~(\ref{relspiral}) and to numerically test their applicability.

\section{Summary}\label{secsum}

In this paper we investigated the relation between the structural changes in the system, represented by the culling time $\tau_{c}$, and between the relaxation time of the persistence function $\tau_{p}$ in the Fredrickson-Andersen and Kob-Andersen kinetically constrained models in one and two dimensions, and in the two-dimensional spiral model. We found that $\tau_{p}\sim\tau^{\gamma}_{c}$, up to logarithmic corrections in the Fredrickson-Andersen and Kob-Andersen models in two dimensions, where $\gamma$ is model-dependent. This result is explained by mapping the persistence of a site to a first passage time of diffusing defects, with their initial distance given by the culling time.

We also found that the persistence function in the $m=2$ models at long times behaves as a stretched exponential $\exp\left[-\left(t/t_{0}\right)^\beta\right]$, where $\beta$ probably goes to zero at small vacancy densities in contradiction to previous studies in which $\beta$ was believed to converge to a finite value. The difference arises because the previous results were obtained in systems which were not in equilibrium, while our simulations are performed in equilibrium.

The general relation between the culling and the persistence times may also hold in other models, including continuum models, and in experiments. Since the $m=1$ Fredrickson-Andersen and Kob-Andersen models represent normal gas or liquid, while the $m=2$ models and the spiral model represent glassy behavior, the exponent $\gamma$ is a measure for the ``glassiness" of a system. It would be interesting to check whether this relation holds also for other measures of the relaxation, such as the autocorrelation function, with the same value of $\gamma$. Also, it would be interesting to study the relation between the culling and the persistence in the three-dimensional extension of the spiral model \cite{spiral3d}, in which there is a decoupling between the structure and the dynamics, namely the density above which permanently jammed structures appear is lower than the density above which the long time self-diffusion stops.

\section*{Acknowledgements}
We thank Roman Golkov, Fabio Leoni, David Mukamel, Nimrod Segall, and Cristina Toninelli for helpful discussions. This research was supported by the Israel Science Foundation grants No. $617/12$, $1730/12$.

\appendix

\section{Exact expression for $P(t)$ in the one-dimensional $m=1$ KA model}

Here we derive an exact expression for the persistence function in the one-dimensional $m=1$ KA model, using the approximation of non-interacting diffusing vacancies. Consider a one-dimensional lattice of length $2L+1$ with $v \cdot (2L+1)$ vacancies. We are interested in the limit $L\rightarrow\infty$, and implicitly take this limit during the derivation whenever there is no singularity. The persistence function, $P(t)$, is the probability that none of these diffusing vacancies reached the origin until time $t$. Since we assume that the vacancies are non-interacting, it is enough to compute the probability that a single vacancy did not reach the origin until time $t$, $Q(t)$, and from that we can obtain $P(t)=\left[Q(t)\right]^{v\cdot (2L+1)}$.

At each time step $\delta$, there is a probability $\delta$ that the vacancy tries to move, and if it does there is an equal probability to move either to the left or to the right. Without loss of generality, we may assume that this vacancy is at site $k_{0}>0$ at time $t=0$. The evolution equation for the probability $v_{k}(t)$ that at time $t$ the vacancy was at site $k$ reads
\begin{align}
v_{k}(t+\delta)=\left(1-\delta\right)v_{k}(t)+\frac{\delta}{2}\left[v_{k+1}(t)+v_{k-1}(t)\right] ,
\end{align}
for $k\geq2$, and
\begin{align}
v_{1}(t+\delta)=\left(1-\delta\right)v_{1}(t)+\frac{\delta}{2}v_{2}(t) ,
\end{align}
for $k=1$, since if the vacancy reached the site $k=0$, the process stops. In the limit $\delta\rightarrow0$ this transforms to the differential equations
\begin{align}
&\frac{dv_{k}}{dt}=-v_{k}+\frac{1}{2}\left(v_{k+1}+v_{k-1}\right) ,\nonumber\\
&\frac{dv_{1}}{dt}=-v_{1}+\frac{1}{2}v_{2} .
\end{align}
The general solution to the first differential equation is \cite{glauber}
\begin{align}
v_{k}=e^{-t}\sum^{\infty}_{l=-\infty}A_{l}I_{l-k}(t) ,
\end{align}
where $I_{n}(t)$ is the modified Bessel function of the first kind. Setting the general solution in the equation for $k=1$ yields
\begin{align}
\sum^{\infty}_{l=-\infty}A_{l}I_{l-k}(t)=0 .
\end{align}
Using the relation $I_{n}(t)=I_{-n}(t)$, we find that
\begin{align}
A_{l}=-A_{-l} .
\end{align}
Imposing the initial condition $v_{k}(0)=\delta_{k,k_{0}}$ and using $I_{n}(0)=\delta_{n,0}$ yields
\begin{align}
A_{k}=\delta_{k,k_{0}}-\delta_{k,-k_{0}} .
\end{align}
Hence,
\begin{align}
v_{k}(t)=e^{-t}\left[I_{k-k_{0}}(t)-I_{k+k_{0}}(t)\right] .
\end{align}

The probability that the vacancy did not reach the origin until time $t$, given that it started from $k_{0}$ is
\begin{align}
Q_{1}(k_{0},t)=\sum^{\infty}_{k=1}v_{k}(t)=e^{-t}\left[I_{k_{0}}(t)+I_{0}(t)+2\sum^{k_{0}-1}_{k=1}I_{k}(t)\right] .
\end{align}
Averaging over all initial states yields
\begin{align}
&Q(t)=\frac{1}{L}\sum^{L}_{k_{0}=1}Q_{1}(k_{0},t)=\nonumber\\
&=\frac{1}{L}\left[\frac{1}{2}-\frac{e^{-t}I_{0}(t)}{2}+Le^{-t}I_{0}(t)+2e^{-t}\sum^{L}_{k_{0}=1}\sum^{k_{0}-1}_{k=1}I_{k}(t)\right] ,\label{eqq1}
\end{align}
where we used
\begin{align}
\sum^{\infty}_{n=1}I_{n}(t)=\frac{e^{t}}{2}-\frac{I_{0}(t)}{2} .\label{bes1}
\end{align}
In order to calculate the last sum, we change the order of summation such that
\begin{align}
&\sum^{L}_{k_{0}=1}\sum^{k_{0}-1}_{k=1}I_{k}(t)=\sum^{\infty}_{k=1}\sum^{L}_{k_{0}=k+1}I_{k}(t)=\nonumber\\
&=\sum^{\infty}_{k=1}\left(L-k\right)I_{k}(t)=\frac{L}{2}\left(e^{t}-I_{0}\right)-\sum^{\infty}_{k=1}kI_{k}(t) ,\label{eqq2}
\end{align}
where we used Eq. (\ref{bes1}). 

We now use the relation
\begin{align}
kI_{k}(t)=\frac{t}{2}\left[I_{k-1}(t)-I_{k+1}(t)\right] ,
\end{align}
and find that
\begin{align}
\sum^{\infty}_{k=1}kI_{k}(t)=\frac{t}{2}\left[I_{0}(t)+I_{1}(t)\right] .\label{eqq3}
\end{align}
Therefore, by combining Eqs. (\ref{eqq1}), (\ref{eqq2}) and (\ref{eqq3}), we find that
\begin{align}
Q(t)=1+\frac{1}{2L}\left[1-e^{-t}I_{0}(t)-te^{-t}\left(I_{0}(t)+I_{1}(t)\right)\right] ,
\end{align}
and thus the persistence function is given by
\begin{align}
&P(t)=\lim_{L\rightarrow\infty}\left[Q(t)\right]^{v(2L+1)}=\nonumber\\
&=\exp\left\{-v\left[e^{-t}I_{0}(t)-1+te^{-t}\left(I_{0}(t)+I_{1}(t)\right)\right]\right\} .
\end{align}

For $t\ll1$ the persistence function behaves as $\exp\left(-vt\right)$, and for $t\gg1$ it behaves as
\begin{align}
P(t\gg1)\approx\exp\left(-v\sqrt{\frac{2t}{\pi}}\right) .
\end{align}

\end{document}